\documentclass[a4paper,11pt]{article}
\usepackage{amsfonts}
\usepackage{amsmath}
\usepackage{amssymb}
\usepackage{geometry}

\input{tcilatex}

\begin{document}

\title{Obtaining consistent Lorentz gauging for a gravitationally coupled
fermion}
\author{John Fredsted\thanks{%
physics@johnfredsted.dk} \\
R\o m\o v\ae nget 32B, 8381 Tilst, Denmark}
\maketitle

\begin{abstract}
For internal gauge forces, the result of locally gauging, i.e., of
performing the substitution $\partial \rightarrow D$, is physically the same
whether performed on the action or on the corresponding Euler-Lagrange
equations of motion. Rather unsettling, though, such commutativity fails for
the standard way of coupling a Dirac fermion to the gravitational field in
the setting of a local Lorentz gauge theory of general relativity in the
vierbein formalism, the equivalence principle thus seemingly being here
violated. This paper will present a formalism in which commutativity holds
for the gravitational force as well, the action for the gravitational field
itself being still the Einstein-Hilbert one. Notably, in this formalism, the
spinor field will carry a world/coordinate index, rather than a Lorentz
spinor index as it does standardly. More generally, no Lorentz indices will
figure, neither vector indices nor spinor indices, which from a parsimonious
point of view seems quite satisfactory.
\end{abstract}

\section{Introduction}

Consider in global Minkowski spacetime, with metric $\eta _{ab}$ in
Euclidian coordinates $x^{a}$, the following free Dirac action (written in
explicitly hermitian form):%
\begin{eqnarray*}
S_{free} &=&\int \mathcal{L}_{free}\sqrt{-\eta }d^{4}x, \\
\mathcal{L}_{free} &=&\frac{\mathrm{i}}{2}\left[ \overline{\psi }\gamma
^{a}\left( \partial _{a}\psi \right) -\overline{\left( \partial _{a}\psi
\right) }\gamma ^{a}\psi \right] -m\overline{\psi }\psi .
\end{eqnarray*}%
Here, the factor $\sqrt{-\eta }=1$, although trivial, has been included for
completeness. The corresponding Euler-Lagrange equations of motion are given
by $E_{free}\equiv \left( \mathrm{i}\gamma ^{a}\partial _{a}-m\right) \psi
=0 $. Let $\psi $ have an electric charge $q$, say. Then in the presence of
an external electromagnetic field $A_{a}$, the Lagrangian is augmented to%
\begin{eqnarray*}
S &=&\int \mathcal{L}\sqrt{-\eta }d^{4}x, \\
\mathcal{L} &=&\frac{\mathrm{i}}{2}\left[ \overline{\psi }\gamma ^{a}\left(
D_{a}\psi \right) -\overline{\left( D_{a}\psi \right) }\gamma ^{a}\psi %
\right] -m\overline{\psi }\psi ,
\end{eqnarray*}%
where $D_{a}=\partial _{a}+\mathrm{i}qA_{\mu }$. The corresponding
Euler-Lagrange equations of motion are now given by $E\equiv \left( \mathrm{i%
}\gamma ^{a}D_{a}-m\right) \psi =0$. Reassuringly, $E$ results whether the
substitution $\partial _{a}\rightarrow D_{a}$ is performed on $S_{free}$ or
on $E_{free}$; the substitution procedure $\partial _{a}\rightarrow D_{a}$
may thus be said to commute with the Euler-Lagrange variational procedure.
This commutativity property holds not only for an electromagnetically
coupled fermion; it holds as well for a weakly coupled fermionic doublet,
and for a strongly coupled fermionic triplet, the reason being that the
generators for the weak and strong forces, respectively, commute with $%
1_{2}\otimes \gamma ^{a}$ and $1_{3}\otimes \gamma ^{a}$, where $1_{n}$
means the $n\times n$ unit matrix. Generally, it holds for any internal
gauge force with generators commuting with $1_{n}\otimes \gamma ^{a}$ (for
appropiate values of $n$).

All this would be pretty uninteresting, though, was it not for the following
fact: such commutativity fails for a \textit{standardly} gravitationally
coupled fermion. Proof: To switch on gravitational and/or inertial forces,
in the realm of general relativity recasted as a local Lorentz gauge theory,
the standard procedure, compare \cite[Sec. 31.A]{Weinberg: QFT} and \cite[%
Sec. 12.1]{GSW}, is 1.) to introduce a vierbein, $e^{\mu }{}_{a}$, and an
associated minimal (i.e., torsionless) spin connection, $\omega _{\mu
}{}^{a}{}_{b}\equiv e^{a}{}_{\rho }\nabla _{\mu }e^{\rho }{}_{b}$, and 2.)
to perform in conjunction the substitutions $\eta _{ab}\rightarrow g_{\mu
\nu }=\eta _{ab}e^{a}{}_{\mu }e^{b}{}_{\nu }$ and $\gamma ^{a}\rightarrow
\gamma ^{\mu }=e^{\mu }{}_{a}\gamma ^{a}$ and%
\begin{equation}
\partial _{\mu }\psi \rightarrow D_{\mu }\psi \equiv \left( \partial _{\mu }+%
\frac{1}{2}\omega _{\mu ab}S^{ab}\right) \psi ,  \label{Eq:LorentzCovDer}
\end{equation}%
where $S^{ab}\equiv \frac{1}{4}\left[ \gamma ^{a},\gamma ^{b}\right] $ are
the generators of the spinor representation of the Lorentz group, being here
defined in the 'mathematicians way' without an explicit $\mathrm{i}$.
Applied to the free action $S_{free}$ previously given, the result is%
\begin{eqnarray*}
S_{grav} &=&\int \mathcal{L}_{grav}\sqrt{-g}d^{4}x, \\
\mathcal{L}_{grav} &=&\frac{\mathrm{i}}{2}e^{\mu }{}_{a}\left[ \overline{%
\psi }\gamma ^{a}\left( D_{\mu }\psi \right) -\overline{\left( D_{\mu }\psi
\right) }\gamma ^{a}\psi \right] -m\overline{\psi }\psi
\end{eqnarray*}%
for a Dirac fermion in an external gravitational field. The corresponding
Euler-Lagrange equations of motion are given by%
\begin{eqnarray*}
0 &=&E_{grav} \\
&\equiv &\left( \mathrm{i}e^{\mu }{}_{a}\gamma ^{a}\partial _{\mu }-m\right)
\psi +\frac{\mathrm{i}}{2}\left( \nabla _{\mu }e^{\mu }{}_{a}\right) \gamma
^{a}\psi +\frac{\mathrm{i}}{4}e^{\mu }{}_{a}\omega _{\mu cd}\left\{ \gamma
^{a},S^{cd}\right\} \psi \\
&=&\left( \mathrm{i}e^{\mu }{}_{a}\gamma ^{a}\partial _{\mu }-m\right) \psi -%
\frac{\mathrm{i}}{2}e^{\mu }{}_{b}\omega _{\mu }{}^{b}{}_{a}\gamma ^{a}\psi +%
\frac{\mathrm{i}}{4}e^{\mu }{}_{a}\omega _{\mu cd}\left\{ \gamma
^{a},S^{cd}\right\} \psi ,
\end{eqnarray*}%
using 1.) the identity $\partial _{\mu }\left( e^{\mu }{}_{a}\sqrt{-g}%
\right) =\nabla _{\mu }e^{\mu }{}_{a}\equiv \partial _{\mu }e^{\mu
}{}_{a}+\Gamma ^{\mu }{}_{\rho \mu }e^{\rho }{}_{a}$, where $\Gamma ^{\mu
}{}_{\nu \rho }$ is the Levi-Civita connection, and 2.) the minimality of
the spin connection. But if the substitution is instead applied to the
Euler-Lagrange equations of motion $E_{free}$ previously given, the result is%
\begin{eqnarray*}
0 &=&\tilde{E}_{grav} \\
&\equiv &\left( \mathrm{i}e^{\mu }{}_{a}\gamma ^{a}\partial _{\mu }-m\right)
\psi +\frac{\mathrm{i}}{2}e^{\mu }{}_{a}\omega _{\mu cd}\gamma ^{a}S^{cd}\psi
\\
&\neq &E_{grav}.
\end{eqnarray*}%
End of proof. This nonequality of $E_{grav}$ and $\tilde{E}_{grav}$ seems to
the author rather unsettling as it seems to imply that the equivalence
principle is violated (for a Dirac fermion): Whereas $\tilde{E}_{grav}$
seems to be the correct way of implementing the equivalence principle, $%
E_{grav}\neq \tilde{E}_{grav}$ derived from an action by Euler-Lagrange
variation must necessarily take precedence over it, thus resulting in an
inconsistency.

The main purpose of this paper is to present a formalism for the coupling of
a fermion to the gravitational field in which no such ambiguity arises,
i.e., in which the substitution procedure $\partial \rightarrow D$, now with
a different $D$ of course, commute with the Euler-Lagrange variational
procedure. The formalism will contain only world indices, with neither
Lorentz vector indices nor Lorentz spinor indices figuring; contrary to what
appears to be standard wisdom, it will prove possible to have the spinor
field carry a world index rather than a Lorentz spinor index.

\section{\label{Sec:PreliminariesGeometry}Preliminaries, I: Geometry}

Let $\left( M,g,\Gamma \right) $ be a Riemannian manifold $M$ equipped with
a metric $g$ of signature $\left( 1,3\right) $ and corresponding Levi-Civita
connection $\Gamma $. Introduce on this manifold one timelike- and three
spacelike vector fields, $n^{\mu }$ and $n_{i}^{\mu }$, respectively,
subject to the following conditions: 
\begin{subequations}
\label{Eq:LorentzFrame}
\begin{eqnarray}
1 &=&g_{\mu \nu }n^{\mu }n^{\nu }=n^{\mu }n_{\mu },
\label{Eq:LorentzFrame_00} \\
0 &=&g_{\mu \nu }n^{\mu }n_{i}^{\nu }=n^{\mu }n_{i\mu },
\label{Eq:LorentzFrame_0i} \\
-\delta _{ij} &=&g_{\mu \nu }n_{i}^{\mu }n_{j}^{\nu }=n_{i}^{\mu }n_{j\mu },
\label{Eq:LorentzFrame_ij}
\end{eqnarray}%
where $n_{\mu }\equiv g_{\mu \nu }n^{\nu }$ and $n_{i\mu }\equiv g_{\mu \nu
}n_{i}^{\nu }$, of course. In conjunction, these four vector fields
constitute a local Lorentz frame. Although thus effectively constituting a
standard vierbein $e^{\mu }{}_{a}$, performing the obvious identifications $%
e^{\mu }{}_{0}=n^{\mu }$ and $e^{\mu }{}_{i}=n_{i}^{\mu }$, no vierbein will
be used in order to avoid introducing what will turn out to be unnecessary
Lorentz (vector) indices. Note that due to Sylvester's law of inertia \cite[%
p. 86]{Weinberg: Gravitation}, the concept of one timelike- and three
spacelike vector fields is a geometrical one: no coordinate transformation
can change the signature. Thus it makes sense, and is quite natural, to $1+3$
decompose the standard vierbein into two sets of vector fields: the single
timelike one $n^{\mu }$, and the three spacelike ones $n_{i}^{\mu }$. The
metric may be expressed as 
\end{subequations}
\begin{eqnarray}
g_{\mu \nu } &=&n_{\mu }n_{\nu }-\delta ^{ij}n_{i\mu }n_{j\nu }  \notag \\
&=&n_{\mu }n_{\nu }-\overline{n}_{\mu }\cdot \overline{n}_{\nu },
\label{Eq:MetricDef}
\end{eqnarray}%
introducing the three-vector of four-vectors $\overline{n}_{\mu }$ by $%
\left( \overline{n}_{\mu }\right) _{i}\equiv n_{i\mu }$, the dot product
being performed over the Latin indices. Here, and below, a bar will denote a
three-vector quantity. As for the standard vierbein, this expression for the
metric in terms of four vector fields introduces excess local degrees of
freedom: the metric is invariant under the following local Lorentz
transformations: 
\begin{subequations}
\begin{eqnarray}
\delta n^{\mu } &=&\frac{1}{2}\left( d\theta _{\alpha \beta }\right) \left( 
\mathbf{V}^{\alpha \beta }\right) ^{\mu }{}_{\nu }n^{\nu }\equiv \left(
d\theta ^{\mu }{}_{\nu }\right) n^{\nu },  \label{Eq:nLorentzTransWorld} \\
\delta \overline{n}^{\mu } &=&\frac{1}{2}\left( d\theta _{\alpha \beta
}\right) \left( \mathbf{V}^{\alpha \beta }\right) ^{\mu }{}_{\nu }\overline{n%
}^{\nu }\equiv \left( d\theta ^{\mu }{}_{\nu }\right) \overline{n}^{\nu },
\label{Eq:nVecLorentzTransWorld}
\end{eqnarray}%
where $d\theta _{\alpha \beta }=-d\theta _{\beta \alpha }\in \mathbb{R}$,
and where the $4\times 4$ \textit{matrices} $\mathbf{V}^{\mu \nu }=-\mathbf{V%
}^{\nu \mu }$ (here and below matrices are set in boldface) with components $%
\left( \mathbf{V}^{\mu \nu }\right) ^{\rho }{}_{\sigma }\equiv g^{\mu \rho
}\delta _{\sigma }^{\nu }-g^{\nu \rho }\delta _{\sigma }^{\mu }$ constitute
the vector representation of the Lorentz algebra in the sense that they
satisfy 
\end{subequations}
\begin{equation}
-\left[ \mathbf{V}^{\mu \nu },\mathbf{V}^{\rho \sigma }\right] =g^{\mu \rho }%
\mathbf{V}^{\nu \sigma }-g^{\mu \sigma }\mathbf{V}^{\nu \rho }-g^{\nu \rho }%
\mathbf{V}^{\mu \sigma }+g^{\nu \sigma }\mathbf{V}^{\mu \rho },
\label{Eq:LorentzVectorRep}
\end{equation}%
i.e., they are the generators of the vector representation of the Lorentz
group.

A remark: Strictly speaking, the above transformation is not a local Lorentz
transformation, as it operates on world indices, rather than on Lorentz
vector indices. But it may, nonetheless, by a mild abuse of terminology
(which will be adhered to in the rest of the paper), be called so for the
following reason: A genuine (infinitesimal) local Lorentz transformation, 
\textit{not} acting on any world indices, is given by 
\begin{subequations}
\begin{eqnarray}
\delta n^{\mu } &=&\overline{d\xi }\cdot \overline{n}^{\mu },
\label{Eq:nLorentzTrans} \\
\delta \overline{n}^{\mu } &=&\left( \overline{d\xi }\right) n^{\mu }+%
\overline{d\theta }\times \overline{n}^{\mu },  \label{Eq:nVecLorentzTrans}
\end{eqnarray}%
where $\overline{d\theta },\overline{d\xi }\in \mathbb{R}^{3}$ are
(spacetime-dependent) infinitesimal rotation and boost parameters,
respectively. However, using Eqs. (\ref{Eq:LorentzFrame_00})-(\ref%
{Eq:LorentzFrame_ij}), they are readily seen to be equal to Eqs. (\ref%
{Eq:nLorentzTransWorld})-(\ref{Eq:nVecLorentzTransWorld}) if the following
one-to-one identification (of the degrees of freedom) is made: 
\end{subequations}
\begin{equation*}
-d\theta _{\alpha \beta }=\left( \overline{n}_{\alpha }\times \overline{n}%
_{\beta }\right) \cdot \overline{d\theta }+\left( n_{\alpha }\overline{n}%
_{\beta }-n_{\beta }\overline{n}_{\alpha }\right) \cdot \overline{d\xi }.
\end{equation*}%
The overall minus sign in this relationship is due to the fact that the
Lorentz transformations of Eqs. (\ref{Eq:nLorentzTransWorld})-(\ref%
{Eq:nVecLorentzTransWorld}) act on \textit{contravariant} world indices,
whereas the Lorentz transformations of Eqs. (\ref{Eq:nLorentzTrans})-(\ref%
{Eq:nVecLorentzTrans}) act on \textit{covariant} Lorentz indices (the $a$ of 
$e^{\mu }{}_{a}$, remembering the previously mentioned possible
identifications $e^{\mu }{}_{0}=n^{\mu }$ and $e^{\mu }{}_{i}=n_{i}^{\mu }$%
). End of remark.

As for the standard vierbein formulation of general relativity, compare
again \cite[Sec. 31.A]{Weinberg: QFT} and \cite[Sec. 12.1]{GSW}, these
excess Lorentz degrees of freedom should be killed in order to avoid
augmenting the standard content of general relativity. As standardly, this
is done by requirering that the local Lorentz frame field consisting of $%
n^{\mu }$ and $\overline{n}^{\mu }$ in conjunction be covariantly constant:%
\begin{eqnarray}
0 &=&D_{\rho }n^{\mu }\equiv \nabla _{\rho }n^{\mu }+\omega ^{\mu }{}_{\nu
\rho }n^{\nu }=\nabla _{\rho }n^{\mu }+\frac{1}{2}\omega _{\alpha \beta \rho
}\left( \mathbf{V}^{\alpha \beta }\right) ^{\mu }{}_{\nu }n^{\nu },
\label{Eq:nCovDer} \\
\overline{0} &=&D_{\rho }\overline{n}^{\mu }\equiv \nabla _{\rho }\overline{n%
}^{\mu }+\omega ^{\mu }{}_{\nu \rho }\overline{n}^{\nu }=\nabla _{\rho }%
\overline{n}^{\mu }+\frac{1}{2}\omega _{\alpha \beta \rho }\left( \mathbf{V}%
^{\alpha \beta }\right) ^{\mu }{}_{\nu }\overline{n}^{\nu },
\label{Eq:nVecCovDer}
\end{eqnarray}%
for some connection $\omega ^{\mu }{}_{\nu \rho }$ to be introduced. The
unique solution to these conditions is 
\begin{subequations}
\label{Eq:SpinConn}
\begin{eqnarray}
\omega ^{\mu }{}_{\nu \rho } &=&n^{\mu }\nabla _{\rho }n_{\nu }-\overline{n}%
^{\mu }\cdot \nabla _{\rho }\overline{n}_{\nu }  \label{Eq:SpinConnVersionI}
\\
&\equiv &-\left[ \left( \nabla _{\rho }n^{\mu }\right) n_{\nu }-\left(
\nabla _{\rho }\overline{n}^{\mu }\right) \cdot \overline{n}_{\nu }\right] ,
\label{Eq:SpinConnVersionII}
\end{eqnarray}%
the identity (of the second line) following from $\delta _{\nu }^{\mu
}=n^{\mu }n_{\nu }-\overline{n}^{\mu }\cdot \overline{n}_{\nu }$. This
connection will in the present formalism be the analogue of the standard
spin connection $\omega _{\mu }{}^{a}{}_{b}$ (note the different ordering of
indices, though: $\mu $ in $\omega _{\mu }{}^{a}{}_{b}$ will correspond to $%
\rho $ in $\omega ^{\mu }{}_{\nu \rho }$). Using Eq. (\ref{Eq:MetricDef}),
this spin connection is readily seen to be metric compatible, $D_{\rho
}g_{\mu \nu }=0$. Under the local Lorentz transformation of Eqs. (\ref%
{Eq:nLorentzTransWorld})-(\ref{Eq:nVecLorentzTransWorld}), it transforms as 
\end{subequations}
\begin{equation}
\delta \omega ^{\mu }{}_{\nu \rho }=\left( d\theta ^{\mu }{}_{\sigma
}\right) \omega ^{\sigma }{}_{\nu \rho }-\left( d\theta ^{\sigma }{}_{\nu
}\right) \omega ^{\mu }{}_{\sigma \rho }-\nabla _{\rho }d\theta ^{\mu
}{}_{\nu }.  \label{Eq:LorentzTransSpinConnWorld}
\end{equation}%
[A remark: The presence of the Levi-Civita covariant derivative in Eq. (\ref%
{Eq:LorentzTransSpinConnWorld}) makes $\delta \omega ^{\mu }{}_{\nu \rho }$
a type $\left( 1,2\right) $ world tensor. By a continuing mild abuse of
terminology, compare previous remark, it is also seen to be a type $\left(
1,1\right) $ Lorentz tensor in the indices $\mu \nu $, just as the expression%
\begin{equation*}
\delta \omega _{\rho }{}^{a}{}_{b}=\left( d\theta ^{a}{}_{c}\right) \omega
_{\rho }{}^{c}{}_{b}-\left( d\theta ^{c}{}_{b}\right) \omega _{\rho
}{}^{a}{}_{c}-\partial _{\rho }d\theta ^{a}{}_{b},
\end{equation*}%
for the standard spin connection is a type $\left( 1,1\right) $ Lorentz
tensor in the indices $ab$. End of remark.] It is readily established that%
\begin{eqnarray*}
\delta \left( D_{\rho }n^{\mu }\right) &=&\left( d\theta ^{\mu }{}_{\nu
}\right) D_{\rho }n^{\nu }, \\
\delta \left( D_{\rho }\overline{n}^{\mu }\right) &=&\left( d\theta ^{\mu
}{}_{\nu }\right) D_{\rho }\overline{n}^{\nu },
\end{eqnarray*}%
as is appropiate for a proper covariant derivative. These relations say that 
$D_{\rho }n^{\mu }$ and $D_{\rho }\overline{n}^{\mu }$ are each type $\left(
1,0\right) $ Lorentz tensors (in the index $\mu $). Therefore%
\begin{eqnarray*}
0 &=&D_{\rho }D_{\sigma }n^{\mu }\equiv \nabla _{\rho }D_{\sigma }n^{\mu
}+\omega ^{\mu }{}_{\nu \rho }D_{\sigma }n^{\nu }, \\
\overline{0} &=&D_{\rho }D_{\sigma }\overline{n}^{\mu }\equiv \nabla _{\rho
}D_{\sigma }\overline{n}^{\mu }+\omega ^{\mu }{}_{\nu \rho }D_{\sigma }%
\overline{n}^{\nu },
\end{eqnarray*}%
from which it follows that%
\begin{eqnarray*}
0 &=&\left[ D_{\rho },D_{\sigma }\right] n^{\mu }=\left( R^{\mu }{}_{\nu
\rho \sigma }+\Omega ^{\mu }{}_{\nu \rho \sigma }\right) n^{\nu }, \\
\overline{0} &=&\left[ D_{\rho },D_{\sigma }\right] \overline{n}^{\mu
}=\left( R^{\mu }{}_{\nu \rho \sigma }+\Omega ^{\mu }{}_{\nu \rho \sigma
}\right) \overline{n}^{\nu },
\end{eqnarray*}%
where%
\begin{eqnarray*}
R^{\mu }{}_{\nu \rho \sigma } &=&\partial _{\rho }\Gamma ^{\mu }{}_{\nu
\sigma }-\partial _{\sigma }\Gamma ^{\mu }{}_{\nu \rho }+\Gamma ^{\mu
}{}_{\tau \rho }\Gamma ^{\tau }{}_{\nu \sigma }-\Gamma ^{\mu }{}_{\tau
\sigma }\Gamma ^{\tau }{}_{\nu \rho }, \\
\Omega ^{\mu }{}_{\nu \rho \sigma } &\equiv &\nabla _{\rho }\omega ^{\mu
}{}_{\nu \sigma }-\nabla _{\sigma }\omega ^{\mu }{}_{\nu \rho }+\omega ^{\mu
}{}_{\tau \rho }\omega ^{\tau }{}_{\nu \sigma }-\omega ^{\mu }{}_{\tau
\sigma }\omega ^{\tau }{}_{\nu \rho },
\end{eqnarray*}%
introducing $\Omega ^{\mu }{}_{\nu \rho \sigma }$. Here, $R^{\mu }{}_{\nu
\rho \sigma }$ is of course the standard Riemann curvature tensor in terms
of the Levi-Civita connection. [A remark: Although $R^{\mu }{}_{\nu \rho
\sigma }$ and $\Omega ^{\mu }{}_{\nu \rho \sigma }$ have closely analogous
structure, the following difference should be noted: Whereas the Levi-Civita
symbols $\Gamma ^{\mu }{}_{\nu \rho }$ transform only as a tensor in the
index $\mu $, the spin connection $\omega ^{\mu }{}_{\nu \sigma }$
transforms as a tensor in all its indices. This explains the appearance of
Levi-Civita covariant derivatives in the definition of $\Omega ^{\mu
}{}_{\nu \rho \sigma }$, as opposed to only the partial derivatives in the
expression for $R^{\mu }{}_{\nu \rho \sigma }$. End of remark.] But then%
\begin{eqnarray*}
0 &=&\left( \left[ D_{\rho },D_{\sigma }\right] n^{\mu }\right) n_{\nu
}-\left( \left[ D_{\rho },D_{\sigma }\right] \overline{n}^{\mu }\right)
\cdot \overline{n}_{\nu } \\
&=&\left( R^{\mu }{}_{\tau \rho \sigma }+\Omega ^{\mu }{}_{\tau \rho \sigma
}\right) \left( n^{\tau }n_{\nu }-\overline{n}^{\tau }\cdot \overline{n}%
_{\nu }\right) \\
&=&R^{\mu }{}_{\nu \rho \sigma }+\Omega ^{\mu }{}_{\nu \rho \sigma },
\end{eqnarray*}%
using $n^{\tau }n_{\nu }-\overline{n}^{\tau }\cdot \overline{n}_{\nu
}=\delta _{\nu }^{\tau }$, from which it follows that $\Omega ^{\mu }{}_{\nu
\rho \sigma }=-R^{\mu }{}_{\nu \rho \sigma }$. As the Riemann tensor is
locally Lorentz invariant, because the metric is so, this immediately
implies that $\Omega ^{\mu }{}_{\nu \rho \sigma }$ is as well. The
Einstein-Hilbert action is obviously proportional to $g^{\mu \rho }g^{\nu
\sigma }\Omega _{\mu \nu \rho \sigma }$.

\section{Preliminaries, II: Algebra}

It will prove useful to define a transposition operator $\mathrm{\hat{T}}$,
say, by%
\begin{eqnarray}
\left( \mathbf{V}^{\mathrm{\hat{T}}}\right) _{\mu } &\equiv &g_{\mu \nu
}V^{\nu }\equiv g_{\mu \nu }\left( \mathbf{V}\right) ^{\nu },
\label{Eq:hatTVector} \\
\left( \mathbf{A}^{\mathrm{\hat{T}}}\right) ^{\mu }{}_{\nu } &\equiv &g^{\mu
\rho }g_{\nu \sigma }A^{\sigma }{}_{\rho }\equiv g^{\mu \rho }g_{\nu \sigma
}\left( \mathbf{A}\right) ^{\sigma }{}_{\rho },  \label{Eq:hatTMatrix}
\end{eqnarray}%
for any four-column vector $\mathbf{V}$, and any $4\times 4$ matrix $\mathbf{%
A}$. Its action \textit{would} become that of the standard transposition
operator $\mathrm{T}$ if $g_{\mu \nu }=\mathrm{diag}\left( 1,1,1,1\right) $.
Note that $\mathbf{V}^{\mathrm{\hat{T}}}$ is, as it should be, a row vector,
carrying a lower/covariant index. [A remark: For any matrix, the row index
will always be an upper/contravariant index, and the column index will
always be a lower/covariant index, with matrix multiplication thus being
given by $\left( \mathbf{AB}\right) ^{\rho }{}_{\sigma }=\left( \mathbf{A}%
\right) ^{\rho }{}_{\tau }\left( \mathbf{B}\right) ^{\tau }{}_{\sigma
}=A^{\rho }{}_{\tau }B^{\tau }{}_{\sigma }$, as usual, for any two matrices $%
\mathbf{A},\mathbf{B}$. End of remark.] It is readily proved that it shares
with $\mathrm{T}$ the properties $\left( \mathbf{AB}\right) ^{\mathrm{\hat{T}%
}}=\mathbf{B}^{\mathrm{\hat{T}}}\mathbf{A}^{\mathrm{\hat{T}}}$ and $\left( 
\mathbf{AV}\right) ^{\mathrm{\hat{T}}}=\mathbf{V}^{\mathrm{\hat{T}}}\mathbf{A%
}^{\mathrm{\hat{T}}}$, for any $4\times 4$ matrices $\mathbf{A},\mathbf{B}$,
and any four-column vector $\mathbf{V}$. Naturally associated with $\mathrm{%
\hat{T}}$ is $\hat{\dagger}$ defined by%
\begin{eqnarray}
\mathbf{V}^{\hat{\dagger}} &\equiv &\left( \mathbf{V}^{\mathrm{\hat{T}}%
}\right) ^{\ast }\equiv \left( \mathbf{V}^{\ast }\right) ^{\mathrm{\hat{T}}},
\label{Eq:hatDaggerVector} \\
\mathbf{A}^{\hat{\dagger}} &\equiv &\left( \mathbf{A}^{\mathrm{\hat{T}}%
}\right) ^{\ast }\equiv \left( \mathbf{A}^{\ast }\right) ^{\mathrm{\hat{T}}}.
\label{Eq:hatDaggerMatrix}
\end{eqnarray}%
A matrix $\mathbf{A}$ for which $\mathbf{A}^{\mathrm{\hat{T}}}=\pm \mathbf{A}
$ will be called hat-(anti)symmetric, and a matrix $\mathbf{A}$ for which $%
\mathbf{A}^{\hat{\dagger}}=\pm \mathbf{A}$ will be called
hat-(anti)hermitian.

\subsection{Concerning Klein-Gordon compatibility}

By the notion 'Klein-Gordon compatibility' is generally meant the
requirement that all solutions to some given Euler-Lagrange equations of
motion are on mass-shell. Consider the following tensorial quantities:%
\begin{eqnarray}
M_{\mu \rho \sigma } &\equiv &+\left( g_{\mu \rho }g_{\nu \sigma }+g_{\mu
\sigma }g_{\nu \rho }-g_{\mu \nu }g_{\rho \sigma }-\mathrm{i}\varepsilon
_{\mu \nu \rho \sigma }\right) n^{\nu },  \label{Eq:MDef} \\
N_{i\rho \sigma } &\equiv &-\left( g_{\mu \rho }g_{\nu \sigma }-g_{\mu
\sigma }g_{\nu \rho }-\mathrm{i}\varepsilon _{\mu \nu \rho \sigma }\right)
n^{\mu }n_{i}^{\nu },  \label{Eq:NDef}
\end{eqnarray}%
where $\varepsilon _{\mu \nu \rho \sigma }\equiv \sqrt{-g}\left[ \mu \nu
\rho \sigma \right] $ is the Levi-Civita tensor in the notation of \cite[Eq.
(8.10a)]{MTW}. Note that $N_{i\rho \sigma }$ constitutes \textit{three} rank
two (world) tensors, one for each value of $i$. Define the $4\times 4$
matrices $\mathbf{M}^{\mu }$ and $\mathbf{N}_{i}$ by%
\begin{eqnarray}
\left( \mathbf{M}^{\mu }\right) ^{\rho }{}_{\sigma } &\equiv &M^{\mu \rho
}{}_{\sigma },  \label{Eq:MMatrixDef} \\
\left( \mathbf{N}_{i}\right) ^{\rho }{}_{\sigma } &\equiv &N_{i}{}^{\rho
}{}_{\sigma }.  \label{Eq:NMatrixDef}
\end{eqnarray}%
They satisfy the following algebra (note that there is no complex
conjugation of $\mathbf{M}^{\mu }$ in the second relation):%
\begin{eqnarray}
2g^{\mu \nu }\mathbf{1} &=&\mathbf{M}^{\mu }\mathbf{M}^{\nu \ast }+\mathbf{M}%
^{\nu }\mathbf{M}^{\mu \ast },  \label{Eq:MMAlgebra} \\
\mathbf{0} &=&\mathbf{M}^{\mu }\mathbf{N}_{i}^{\ast }+\mathbf{N}_{i}\mathbf{M%
}^{\mu },  \label{Eq:MNAlgebra}
\end{eqnarray}%
where $\mathbf{1}$ and $\mathbf{0}$ are respectively the $4\times 4$
identity matrix and the $4\times 4$ zero matrix. Furthermore, the matrices $%
\mathbf{N}_{i}$ satisfy the following algebra:%
\begin{equation}
\mathbf{N}_{i}\mathbf{N}_{j}=\delta _{ij}\mathbf{1}\pm \mathrm{i}\varepsilon
_{ijk}\mathbf{N}^{k}.  \label{Eq:NPauliAlgebra}
\end{equation}%
The sign in front of the Levi-Civita symbol depends on whether the three
spacelike $n_{i}^{\mu }$ form a right-handed basis (plus sign) or
left-handed basis (minus sign) when considered as three-vectors in the $3D$
subspace they span. Note that for the right-handed case, $\mathbf{N}_{i}$
thus obey the same algebra as do the Pauli matrices. Eqs. (\ref{Eq:MMAlgebra}%
)-(\ref{Eq:NPauliAlgebra}) are relevant for the proof of Klein-Gordon
compatibility of some Euler-Lagrange equations of motion to be derived in
Sec. \ref{Sec:Action} below. In particular, Eq. (\ref{Eq:MMAlgebra}) will in
the present formalism play a role analogous to the Dirac algebra (of the
gamma matrices) in the standard Dirac algebra. The matrices $\mathbf{M}^{\mu
}$ and $\mathbf{N}_{i}$ are respectively hat-hermitian and hat-antisymmetric:%
\begin{eqnarray}
\mathbf{M}^{\mu \hat{\dagger}} &=&+\mathbf{M}^{\mu },  \label{Eq:hatHermM} \\
\mathbf{N}_{i}^{\mathrm{\hat{T}}} &=&-\mathbf{N}_{i}.  \label{Eq:hatAntiSymN}
\end{eqnarray}%
These two relations are relevant for the proof of hermiticity (complex
self-conjugacy) of the Lagrangian to be studied in Sec. \ref{Sec:Action}
below.

\subsection{Concerning Lorentz invariance}

Introduce the $4\times 4$ matrices $\mathbf{S}^{\mu \nu }=-\mathbf{S}^{\nu
\mu }$ by%
\begin{equation}
4\mathbf{S}^{\mu \nu }\equiv \mathbf{M}^{\mu }\mathbf{M}^{\nu \ast }-\mathbf{%
M}^{\nu }\mathbf{M}^{\mu \ast },  \label{Eq:SDef}
\end{equation}%
where $\mathbf{M}^{\mu }$ is given by Eq. (\ref{Eq:MMatrixDef}). They
satisfy the following relations (note that there is no complex conjugation
of $\mathbf{S}^{\mu \nu }$ in the second relation):%
\begin{eqnarray}
\mathbf{M}^{\rho }\mathbf{S}^{\mu \nu \ast }-\mathbf{S}^{\mu \nu }\mathbf{M}%
^{\rho } &=&\left( \mathbf{V}^{\mu \nu }\right) ^{\rho }{}_{\sigma }\mathbf{M%
}^{\sigma },  \label{Eq:MSAlgebra} \\
\mathbf{N}_{i}\mathbf{S}^{\mu \nu }-\mathbf{S}^{\mu \nu }\mathbf{N}_{i} &=&%
\mathbf{0}.  \label{Eq:NSAlgebra}
\end{eqnarray}%
These relations are readily proved using respectively Eq. (\ref{Eq:MMAlgebra}%
) and Eq. (\ref{Eq:MNAlgebra}). The proof of Eq. (\ref{Eq:MSAlgebra}), in
particular, is structurally analogous to the proof, using the Dirac algebra
of gamma matrices, of the identity $\left[ \gamma ^{c},S^{ab}\right] =\left(
V^{ab}\right) ^{c}{}_{d}\gamma ^{d}$ in the standard Dirac formalism, the
only difference being the appearence of complex conjugations here and there.
These matrices $\mathbf{S}^{\mu \nu }$ constitute the spinor representation
of the Lorentz algebra in the sense that they satisfy%
\begin{equation}
-\left[ \mathbf{S}^{\mu \nu },\mathbf{S}^{\rho \sigma }\right] =g^{\mu \rho }%
\mathbf{S}^{\nu \sigma }-g^{\mu \sigma }\mathbf{S}^{\nu \rho }-g^{\nu \rho }%
\mathbf{S}^{\mu \sigma }+g^{\nu \sigma }\mathbf{S}^{\mu \rho },
\label{Eq:LorentzSpinorRep}
\end{equation}%
i.e., they are the generators of the spinor representation of the Lorentz
group. This algebra is readily proved using Eq. (\ref{Eq:MSAlgebra}), the
proof being structurally analogous to the proof, using $\left[ \gamma
^{c},S^{ab}\right] =\left( V^{ab}\right) ^{c}{}_{d}\gamma ^{d}$, of the fact
that $S^{ab}=\frac{1}{4}\left[ \gamma ^{a},\gamma ^{b}\right] $ in the
standard Dirac formalism constitute the spinor representation of the Lorentz
algebra, the only difference being, as before, the appearence of complex
conjugations here and there. These generators are related to the previously
introduced vector representation $\mathbf{V}^{\mu \nu }$, compare Sec. \ref%
{Sec:PreliminariesGeometry}, as%
\begin{equation}
2\left( \mathbf{S}^{\mu \nu }\right) ^{\rho \sigma }=\left( \mathbf{V}^{\mu
\nu }\right) ^{\rho \sigma }+\mathrm{i}\varepsilon ^{\mu \nu \rho \sigma },
\label{Eq:SVRelation}
\end{equation}%
from which it follows that $\mathbf{S}^{\mu \nu }$ is self-dual, $\mathbf{S}%
^{\mu \nu }=\frac{\mathrm{i}}{2}\varepsilon ^{\mu \nu \rho \sigma }\mathbf{S}%
_{\rho \sigma }$, and that%
\begin{eqnarray}
\mathbf{S}^{\mu \nu }+\mathbf{S}^{\mu \nu \mathrm{\hat{T}}} &=&\mathbf{0},
\label{Eq:SPlusST} \\
\mathbf{S}^{\mu \nu }+\mathbf{S}^{\mu \nu \ast } &=&\mathbf{V}^{\mu \nu },
\label{Eq:SPlusSStar}
\end{eqnarray}%
using $\mathbf{V}^{\mu \nu }+\mathbf{V}^{\mu \nu \mathrm{\hat{T}}}=\mathbf{0}
$ and the identity $g^{\rho \alpha }g_{\sigma \beta }\varepsilon ^{\mu \nu
\beta }{}_{\alpha }=-\varepsilon ^{\mu \nu \rho }{}_{\sigma }$. Using Eq. (%
\ref{Eq:SPlusSStar}), to switch $\mathbf{S}^{\mu \nu }$ into $\mathbf{S}%
^{\mu \nu \ast }$ (plus some $\mathbf{V}^{\mu \nu }$), Eqs. (\ref%
{Eq:MSAlgebra})-(\ref{Eq:NSAlgebra}) may be rewritten as%
\begin{eqnarray*}
\mathbf{S}^{\mu \nu \ast }\mathbf{M}^{\rho }-\mathbf{M}^{\rho }\mathbf{S}%
^{\mu \nu } &=&\left[ \mathbf{V}^{\mu \nu },\mathbf{M}^{\rho }\right]
+\left( \mathbf{V}^{\mu \nu }\right) ^{\rho }{}_{\sigma }\mathbf{M}^{\sigma
}, \\
\mathbf{S}^{\mu \nu \ast }\mathbf{N}_{i}-\mathbf{N}_{i}\mathbf{S}^{\mu \nu
\ast } &=&\left[ \mathbf{V}^{\mu \nu },\mathbf{N}_{i}\right] ,
\end{eqnarray*}%
which using Eq. (\ref{Eq:SPlusST}) and the definition of $\hat{\dagger}$,
Eq. (\ref{Eq:hatDaggerMatrix}), may also be written as%
\begin{eqnarray}
\mathbf{S}^{\mu \nu \hat{\dagger}}\mathbf{M}^{\rho }+\mathbf{M}^{\rho }%
\mathbf{S}^{\mu \nu } &=&-\left[ \mathbf{V}^{\mu \nu },\mathbf{M}^{\rho }%
\right] -\left( \mathbf{V}^{\mu \nu }\right) ^{\rho }{}_{\sigma }\mathbf{M}%
^{\sigma },  \label{Eq:SMLorentz} \\
\mathbf{S}^{\mu \nu \hat{\dagger}}\mathbf{N}_{i}+\mathbf{N}_{i}\mathbf{S}%
^{\mu \nu \ast } &=&-\left[ \mathbf{V}^{\mu \nu },\mathbf{N}_{i}\right] .
\label{Eq:SNLorentz}
\end{eqnarray}%
As the metric, and thus as well the Levi-Civita tensor, is invariant under
local Lorentz transformations of $n^{\mu }$ and $\overline{n}^{\mu }$, Eqs. (%
\ref{Eq:nLorentzTransWorld})-(\ref{Eq:nVecLorentzTransWorld}), these
transformations of $n^{\mu }$ and $\overline{n}^{\mu }$ induce via Eqs. (\ref%
{Eq:MDef})-(\ref{Eq:NDef}) the following relations:%
\begin{eqnarray*}
\delta M_{\mu \rho \sigma } &=&-\left( d\theta \right) ^{\tau }{}_{\mu
}M_{\tau \rho \sigma }-\left( d\theta \right) ^{\tau }{}_{\rho }M_{\mu \tau
\sigma }-\left( d\theta \right) ^{\tau }{}_{\sigma }M_{\mu \rho \tau }, \\
\delta N_{i\rho \sigma } &=&-\left( d\theta \right) ^{\tau }{}_{\rho
}N_{i\tau \sigma }-\left( d\theta \right) ^{\tau }{}_{\sigma }N_{i\rho \tau
},
\end{eqnarray*}%
i.e., $M_{\mu \rho \sigma }$ and $N_{i\rho \sigma }$ transform as type $%
\left( 0,3\right) $ and $\left( 0,2\right) $ Lorentz tensors, respectively;
or, equivalently, by raising various indices appropiately:%
\begin{eqnarray}
\delta M^{\mu \rho }{}_{\sigma } &=&\left( d\theta \right) ^{\rho }{}_{\tau
}M^{\mu \tau }{}_{\sigma }-\left( d\theta \right) ^{\tau }{}_{\sigma }M^{\mu
\rho }{}_{\tau }+\left( d\theta \right) ^{\mu }{}_{\tau }M^{\tau \rho
}{}_{\sigma },  \label{Eq:MLorentzTrans} \\
\delta N_{i}{}^{\rho }{}_{\sigma } &=&\left( d\theta \right) ^{\rho
}{}_{\tau }N_{i}{}^{\tau }{}_{\sigma }-\left( d\theta \right) ^{\tau
}{}_{\sigma }N_{i}{}^{\rho }{}_{\tau },  \label{Eq:NLorentzTrans}
\end{eqnarray}%
i.e., $M^{\mu \rho }{}_{\sigma }$ and $\delta N_{i}{}^{\rho }{}_{\sigma }$
transform as type $\left( 2,1\right) $ and $\left( 1,1\right) $ Lorentz
tensors, respectively. Using the almost trivial identity $d\theta ^{\mu
}{}_{\nu }=\frac{1}{2}\left( d\theta _{\alpha \beta }\right) \left( \mathbf{V%
}^{\alpha \beta }\right) ^{\mu }{}_{\nu }$, they may be written in matrix
form as%
\begin{eqnarray}
\delta \mathbf{M}^{\mu } &=&\frac{1}{2}\left( d\theta _{\alpha \beta
}\right) \left[ \mathbf{V}^{\alpha \beta },\mathbf{M}^{\mu }\right] +\frac{1%
}{2}\left( d\theta _{\alpha \beta }\right) \left( \mathbf{V}^{\alpha \beta
}\right) ^{\mu }{}_{\nu }\mathbf{M}^{\nu },  \label{Eq:MMatrixLorentzTrans}
\\
\delta \mathbf{N}_{i} &=&\frac{1}{2}\left( d\theta _{\alpha \beta }\right) %
\left[ \mathbf{V}^{\alpha \beta },\mathbf{N}_{i}\right] .
\label{Eq:NMatrixLorentzTrans}
\end{eqnarray}%
Eq. (\ref{Eq:MMatrixLorentzTrans}) may be compared with the relation $\delta
\gamma ^{\mu }=\frac{1}{2}\left( d\theta _{\alpha \beta }\right) \left( 
\mathbf{V}^{\alpha \beta }\right) ^{\mu }{}_{\nu }\gamma ^{\nu }$, where $%
\gamma ^{\mu }\equiv e^{\mu }{}_{\alpha }\gamma ^{a}$, in the standard Dirac
(vierbein)formalism. The 'extra' commutator term in Eq. (\ref%
{Eq:MMatrixLorentzTrans}) is due to the fact that the row and column indices
of $\mathbf{M}^{\mu }$ transform as well under Lorentz transformations, in
compliance with Eq. (\ref{Eq:MLorentzTrans}). This is of course also the
reason for the commutator term in Eq. (\ref{Eq:NMatrixLorentzTrans}). In
conjunction with Eqs. (\ref{Eq:SMLorentz})-(\ref{Eq:SNLorentz}), these
relations then finally imply that%
\begin{eqnarray}
\delta \mathbf{M}^{\mu } &=&-\frac{1}{2}\left( d\theta _{\alpha \beta
}\right) \left( \mathbf{S}^{\alpha \beta \hat{\dagger}}\mathbf{M}^{\mu }+%
\mathbf{M}^{\mu }\mathbf{S}^{\alpha \beta }\right) ,
\label{Eq:MMatrixLorentzTransFinal} \\
\delta \mathbf{N}_{i} &=&-\frac{1}{2}\left( d\theta _{\alpha \beta }\right)
\left( \mathbf{S}^{\alpha \beta \hat{\dagger}}\mathbf{N}_{i}+\mathbf{N}_{i}%
\mathbf{S}^{\alpha \beta \ast }\right) .  \label{Eq:NMatrixLorentzTransFinal}
\end{eqnarray}%
These two relations are relevant for the proof of (local) Lorentz invariance
of some specific action to be introduced in Sec. \ref{Sec:Action} below.
From Eqs. (\ref{Eq:MLorentzTrans})-(\ref{Eq:NLorentzTrans}) it follows that
the Lorentz covariant derivatives of $M^{\mu \rho }{}_{\sigma }$ and $%
N_{i}{}^{\rho }{}_{\sigma }$ are necessarily given by%
\begin{eqnarray*}
0 &\equiv &D_{\tau }M^{\mu \rho }{}_{\sigma }\equiv \nabla _{\tau }M^{\mu
\rho }{}_{\sigma }+\omega ^{\rho }{}_{\nu \tau }M^{\mu \nu }{}_{\sigma
}-\omega ^{\nu }{}_{\sigma \tau }M^{\mu \rho }{}_{\nu }+\omega ^{\mu
}{}_{\nu \tau }M^{\nu \rho }{}_{\sigma }, \\
0 &\equiv &D_{\tau }N_{i}{}^{\rho }{}_{\sigma }\equiv \nabla _{\tau
}N_{i}{}^{\rho }{}_{\sigma }+\omega ^{\rho }{}_{\nu \tau }N_{i}{}^{\nu
}{}_{\sigma }-\omega ^{\nu }{}_{\sigma \tau }N_{i}{}^{\rho }{}_{\nu },
\end{eqnarray*}%
the identically vanishing of which is due to Eqs. (\ref{Eq:MetricDef}), (\ref%
{Eq:nCovDer})-(\ref{Eq:nVecCovDer}), and (\ref{Eq:MDef})-(\ref{Eq:NDef}).
Using the almost trivial identity $\omega ^{\rho }{}_{\sigma \mu }=\frac{1}{2%
}\left( \omega _{\alpha \beta \mu }\right) \left( \mathbf{V}^{\alpha \beta
}\right) ^{\rho }{}_{\sigma }$, they may be written in matrix form as%
\begin{eqnarray}
\mathbf{0} &\equiv &\mathbf{D}_{\tau }\mathbf{M}^{\mu }=\mathbf{\nabla }%
_{\tau }\mathbf{M}^{\mu }+\frac{1}{2}\omega _{\alpha \beta \tau }\left[ 
\mathbf{V}^{\alpha \beta },\mathbf{M}^{\mu }\right] +\frac{1}{2}\omega
_{\alpha \beta \tau }\left( \mathbf{V}^{\alpha \beta }\right) ^{\mu }{}_{\nu
}\mathbf{M}^{\nu },  \label{Eq:MCovDer} \\
\mathbf{0} &\equiv &\mathbf{D}_{\tau }\mathbf{N}_{i}=\mathbf{\nabla }_{\tau }%
\mathbf{N}_{i}+\frac{1}{2}\omega _{\alpha \beta \tau }\left[ \mathbf{V}%
^{\alpha \beta },\mathbf{N}_{i}\right] ,  \label{Eq:NCovDer}
\end{eqnarray}%
where $\mathbf{\nabla }_{\tau }\mathbf{M}^{\mu }$ and $\mathbf{\nabla }%
_{\tau }\mathbf{N}_{i}$ (boldfaced nablas) mean, respectively, type $\left(
2,2\right) $ and type $\left( 1,2\right) $ world tensors with components $%
\left( \mathbf{\nabla }_{\tau }\mathbf{M}^{\mu }\right) ^{\rho }{}_{\sigma
}\equiv \nabla _{\tau }M^{\mu \rho }{}_{\sigma }$ and $\left( \mathbf{\nabla 
}_{\tau }\mathbf{N}_{i}\right) ^{\rho }{}_{\sigma }\equiv \nabla _{\tau
}N_{i}{}^{\rho }{}_{\sigma }$. [Notational remark: A boldfaced $\nabla $
and/or $D$ will be used whenever the covariant derivative acts on a
matrix/vector-valued quantity, to remind the reader that the Levi-Civita
covariant derivative will have to act also on the hidden row and/or column 
\textit{world} indices, thus producing one or two extra Christoffel terms
when fully expanded in tensor components. For the case just given, $\nabla
_{\tau }\mathbf{M}^{\mu }$ and $\nabla _{\tau }\mathbf{N}_{i}$ (no boldface)
could be mistaken to mean $\nabla _{\tau }\mathbf{M}^{\mu }=\partial _{\tau }%
\mathbf{M}^{\mu }+\Gamma ^{\mu }{}_{\nu \tau }\mathbf{M}^{\nu }\neq \mathbf{%
\nabla }_{\tau }\mathbf{M}^{\mu }$ and $\nabla _{\tau }\mathbf{N}%
_{i}=\partial _{\tau }\mathbf{N}_{i}\neq \mathbf{\nabla }_{\tau }\mathbf{N}%
_{i}$. End of remark.]

A final note: The vector and spinor representations $\mathbf{V}^{\mu \nu }$
and $\mathbf{S}^{\mu \nu }$, which both depend only on the metric, are both
(locally) Lorentz invariant, i.e., $\delta \mathbf{V}^{\mu \nu }=\delta 
\mathbf{S}^{\mu \nu }=\mathbf{0}$. This is reassuring, as the opposite case,
i.e., having Lorentz generators that were \textit{not} Lorentz invariant,
would be somewhat of a conceptual quagmire.

\section{\label{Sec:Action}Action and Euler-Lagrange equations of motion}

This is the main section of the paper in which the pieces laid out in the
previous section (on preliminaries) come together.

Consider in global Minkowski spacetime, in Cartesian coordinates $x^{\mu }$,
endowed with spacetime-\textit{in}dependent $n^{\mu }$ and $\overline{n}%
^{\mu }$ for which $g_{\mu \nu }\equiv n_{\mu }n_{\nu }-\overline{n}_{\mu
}\cdot \overline{n}_{\nu }=\eta _{\mu \nu }$, the following action:%
\begin{eqnarray*}
S_{free}^{W} &=&\int \mathcal{L}_{free}^{W}\sqrt{-\eta }d^{4}x, \\
\mathcal{L}_{free}^{W} &=&\frac{\mathrm{i}}{2}\left[ \psi ^{\rho \ast
}M^{\mu }{}_{\rho \sigma }\left( \partial _{\mu }\psi ^{\sigma }\right)
-\left( \partial _{\mu }\psi ^{\rho }\right) ^{\ast }M^{\mu }{}_{\rho \sigma
}\psi ^{\sigma }\right] -\frac{m}{2}\left( \psi ^{\rho \ast }N_{\rho \sigma
}\psi ^{\sigma \ast }-\psi ^{\rho }N_{\rho \sigma }^{\ast }\psi ^{\sigma
}\right) \\
&\equiv &\frac{\mathrm{i}}{2}\left[ \psi _{\rho }^{\ast }M^{\mu \rho
}{}_{\sigma }\left( \partial _{\mu }\psi ^{\sigma }\right) -\left( \partial
_{\mu }\psi _{\rho }\right) ^{\ast }M^{\mu \rho }{}_{\sigma }\psi ^{\sigma }%
\right] -\frac{m}{2}\left[ \psi _{\rho }^{\ast }N^{\rho }{}_{\sigma }\psi
^{\sigma \ast }-\psi ^{\rho }\left( N^{\rho }{}_{\sigma }\right) ^{\ast
}\psi ^{\sigma }\right] ,
\end{eqnarray*}%
where $N_{\rho \sigma }=N_{\rho \sigma }\left( a\right) \equiv a^{i}N_{i\rho
\sigma }$ for some constants $a^{i}\in \mathbb{R}$ obeying $a^{i}a_{i}=1$.
Here, $M_{\mu \rho \sigma }$ and $N_{i\rho \sigma }$ are given by Eqs. (\ref%
{Eq:MDef})-(\ref{Eq:NDef}). The action will be considered at the classical
level using (complex) Grassmann-valued $\psi ^{\rho }$. The most distinctive
property of $\mathcal{L}_{free}^{W}$ is that the spinor field carries a
world index (the letter $W$ referring to this), as advertised in the
Introduction, rather than a standard (Lorentz) spinor index. Using Eqs. (\ref%
{Eq:hatDaggerVector})-(\ref{Eq:hatDaggerMatrix}), the Lagrangian may also be
written in matrix notation as%
\begin{equation}
\mathcal{L}_{free}^{W}=\frac{\mathrm{i}}{2}\left[ \mathbf{\psi }^{\hat{%
\dagger}}\mathbf{M}^{\mu }\left( \partial _{\mu }\mathbf{\psi }\right)
-\left( \partial _{\mu }\mathbf{\psi }\right) ^{\hat{\dagger}}\mathbf{M}%
^{\mu }\mathbf{\psi }\right] -\frac{m}{2}\left( \mathbf{\psi }^{\hat{\dagger}%
}\mathbf{N\psi }^{\ast }-\mathbf{\psi }^{\mathrm{\hat{T}}}\mathbf{N}^{\ast }%
\mathbf{\psi }\right) ,  \label{Eq:LFreeWorld}
\end{equation}%
where $\mathbf{M}^{\mu }$ and $\mathbf{N}=a^{i}\mathbf{N}_{i}$ are
determined by Eqs. (\ref{Eq:MMatrixDef})-(\ref{Eq:NMatrixDef}), and where $%
\mathbf{\psi }$ is the four-column vector with components $\left( \mathbf{%
\psi }\right) ^{\mu }=\psi ^{\mu }$, obviously. The hat-hermiticity of $%
\mathbf{M}^{\mu }$, Eq. (\ref{Eq:hatHermM}), guarantees that the kinetic
part of the Lagrangian is complex self-conjugate; and the hat-antisymmetry
of $\mathbf{N}$, Eq. (\ref{Eq:hatAntiSymN}), guarantees that the
Majorana-like mass term is both complex self-conjugate and nontrivial. Eqs. (%
\ref{Eq:MMatrixLorentzTransFinal})-(\ref{Eq:NMatrixLorentzTransFinal})
guarantee that the Lagrangian is \textit{globally} invariant under the
following Lorentz transformation (of the fundamental fields): 
\begin{subequations}
\label{Eq:LorentzTransWorld}
\begin{eqnarray}
\delta n^{\mu } &=&\frac{1}{2}\left( d\theta _{\alpha \beta }\right) \left( 
\mathbf{V}^{\alpha \beta }\right) ^{\mu }{}_{\nu }n^{\nu },
\label{Eq:nLorentzTransWorld_} \\
\delta \overline{n}^{\mu } &=&\frac{1}{2}\left( d\theta _{\alpha \beta
}\right) \left( \mathbf{V}^{\alpha \beta }\right) ^{\mu }{}_{\nu }\overline{n%
}^{\nu },  \label{Eq:nVecLorentzTransWorld_} \\
\delta \psi ^{\mu } &=&\frac{1}{2}\left( d\theta _{\alpha \beta }\right)
\left( \mathbf{S}^{\alpha \beta }\right) ^{\mu }{}_{\nu }\psi ^{\nu }.
\label{Eq:psiLorentzTransWorld}
\end{eqnarray}%
Eqs. (\ref{Eq:nLorentzTransWorld_})-(\ref{Eq:nVecLorentzTransWorld_}) are
one-to-one reiterations of Eqs. (\ref{Eq:nLorentzTransWorld})-(\ref%
{Eq:nVecLorentzTransWorld}); remember that it is these transformations which
induce on $\mathbf{M}^{\mu }$ and $\mathbf{N}_{i}$ the transformations given
by Eqs. (\ref{Eq:MMatrixLorentzTransFinal})-(\ref%
{Eq:NMatrixLorentzTransFinal}). The last relation may also be written
succinctly in matrix notation as $\delta \mathbf{\psi }=\frac{1}{2}\left(
d\theta _{\alpha \beta }\right) \mathbf{S}^{\alpha \beta }\mathbf{\psi }$.
The corresponding Euler-Lagrange equations of motion $E_{free}^{W}$ obtained
by varying $S_{free}^{W}$ with respect to $\psi ^{\rho \ast }$ are readily
found to be 
\end{subequations}
\begin{equation*}
0=E_{free}^{W}\equiv \mathrm{i}M^{\mu }{}_{\rho \sigma }\partial _{\mu }\psi
^{\sigma }-mN_{\rho \sigma }\psi ^{\sigma \ast },
\end{equation*}%
using the spacetime-independency of $M^{\mu }{}_{\rho \sigma }$ and $N_{\rho
\sigma }$, due to the assumed spacetime-independency of $n^{\mu }$ and $%
\overline{n}^{\mu }$; or, equivalently, in matrix notation (by raising the $%
\rho $-index):%
\begin{equation}
\mathbf{0}=\mathbf{E}_{free}^{W}\equiv \mathrm{i}\mathbf{M}^{\mu }\partial
_{\mu }\mathbf{\psi }-m\mathbf{N\psi }^{\ast }.  \label{Eq:EFreeWorld}
\end{equation}%
As they should be, these Euler-Lagrange equations of motion are Klein-Gordon
compatible, i.e., any plane wave solution is on mass shell, because%
\begin{eqnarray*}
0 &=&\left( \mathrm{i}\mathbf{M}^{\mu }K\partial _{\mu }-m\mathbf{N}\right)
\left( \mathrm{i}\mathbf{M}^{\nu }K\partial _{\nu }-m\mathbf{N}\right) \\
&=&\mathbf{M}^{\mu }\mathbf{M}^{\nu \ast }\partial _{\mu }\partial _{\nu }-%
\mathrm{i}m\left( \mathbf{M}^{\mu }\mathbf{N}^{\ast }+\mathbf{NM}^{\mu
}\right) K\partial _{\mu }+m^{2}\mathbf{N}^{2} \\
&=&\left( \square +m^{2}\right) \mathbf{1},
\end{eqnarray*}%
using 1.) $\mathbf{E}_{free}^{W}$ in the form $\mathbf{E}_{free}^{W}=\left( 
\mathrm{i}\mathbf{M}^{\mu }K\partial _{\mu }-m\mathbf{N}\right) K\mathbf{%
\psi }$ with $K$ the operator of complex conjugation, 2.) Eqs. (\ref%
{Eq:MMAlgebra})-(\ref{Eq:MNAlgebra}), and 3.) $\mathbf{N}^{2}=a^{i}a^{j}%
\mathbf{N}_{i}\mathbf{N}_{j}=\mathbf{1}$ due to Eq. (\ref{Eq:NPauliAlgebra})
and the constraint $a^{i}a_{i}=1$, compare previous.

Now, in analogy with the standard procedure for switching on gravitational
and/or inertial forces, compare the Introduction, 1.) let the coordinates be
arbitrary, 2.) let $n^{\mu }$ and $\overline{n}^{\mu }$ be subject only to
the orthonormality conditions given by Eq. (\ref{Eq:LorentzFrame}), the
metric $g_{\mu \nu }$ itself thus becoming arbitrary, and 3.) perform the
substitution%
\begin{equation}
\partial _{\mu }\psi ^{\rho }\rightarrow D_{\mu }\psi ^{\rho }\equiv \left[
\delta _{\sigma }^{\rho }\nabla _{\mu }+\frac{1}{2}\omega _{\alpha \beta \mu
}\left( \mathbf{S}^{\alpha \beta }\right) ^{\rho }{}_{\sigma }\right] \psi
^{\sigma },  \label{Eq:LorentzCovDerWorld}
\end{equation}%
with $\omega _{\alpha \beta \mu }$ determined by Eq. (\ref{Eq:SpinConn}),
and $\mathbf{S}^{\alpha \beta }$ determined by Eq. (\ref{Eq:SDef}). Note
that as the spinor field now carries a world index, the explicit appearence
of $\nabla _{\mu }$ is mandatory, in contrast to Eq. (\ref{Eq:LorentzCovDer}%
), where $\psi $ carries only a spinor index. In matrix notation, Eq. (\ref%
{Eq:LorentzCovDerWorld}) may be written as%
\begin{equation}
\partial _{\mu }\mathbf{\psi }\rightarrow \mathbf{D}_{\mu }\mathbf{\psi }%
\equiv \left( \mathbf{\nabla }_{\mu }+\frac{1}{2}\omega _{\alpha \beta \mu }%
\mathbf{S}^{\alpha \beta }\right) \mathbf{\psi },
\label{Eq:LorentzCovDerWorldMatrix}
\end{equation}%
where $\mathbf{\nabla }_{\mu }\mathbf{\psi }$ means a type $\left(
1,1\right) $ world tensor field with components $\left( \mathbf{\nabla }%
_{\mu }\mathbf{\psi }\right) ^{\rho }=\nabla _{\mu }\psi ^{\rho }$, compare
previous remark concerning boldfaced nabla. As shown in the Appendix at the
end of this paper, $\mathbf{D}_{\mu }\mathbf{\psi }$ is a proper Lorentz
covariant derivative in the sense that it transforms as $\delta \left( 
\mathbf{D}_{\mu }\mathbf{\psi }\right) =\frac{1}{2}\left( d\theta _{\alpha
\beta }\right) \mathbf{S}^{\alpha \beta }\mathbf{D}_{\mu }\mathbf{\psi }$
under Eq. (\ref{Eq:LorentzTransWorld}) with $d\theta _{a\beta }$ arbitrary.
The action $S_{free}^{W}$ above then becomes the following coordinate
invariant and \textit{locally} Lorentz invariant action:%
\begin{eqnarray*}
S_{grav}^{W} &=&\int \mathcal{L}_{grav}^{W}\sqrt{-g}d^{4}x, \\
\mathcal{L}_{grav}^{W} &=&\frac{\mathrm{i}}{2}\left[ \psi ^{\rho \ast
}M^{\mu }{}_{\rho \sigma }\left( D_{\mu }\psi ^{\sigma }\right) -\left(
D_{\mu }\psi ^{\rho }\right) ^{\ast }M^{\mu }{}_{\rho \sigma }\psi ^{\sigma }%
\right] -\frac{m}{2}\left( \psi ^{\rho \ast }N_{\rho \sigma }\psi ^{\sigma
\ast }-\psi ^{\rho }N_{\rho \sigma }^{\ast }\psi ^{\sigma }\right) \\
&\equiv &\frac{\mathrm{i}}{2}\left[ \psi _{\rho }^{\ast }M^{\mu \rho
}{}_{\sigma }\left( D_{\mu }\psi ^{\sigma }\right) -\left( D_{\mu }\psi
_{\rho }\right) ^{\ast }M^{\mu \rho }{}_{\sigma }\psi ^{\sigma }\right] -%
\frac{m}{2}\left[ \psi _{\rho }^{\ast }N^{\rho }{}_{\sigma }\psi ^{\sigma
\ast }-\psi ^{\rho }\left( N^{\rho }{}_{\sigma }\right) ^{\ast }\psi
^{\sigma }\right] ,
\end{eqnarray*}%
where now $M^{\mu }{}_{\rho \sigma }$ and $N_{\rho \sigma }$ are generally
spacetime-dependent. Explicitly expanding the covariant derivatives, the
Lagrangian is given by%
\begin{eqnarray*}
\mathcal{L}_{grav}^{W} &=&\frac{\mathrm{i}}{2}\psi ^{\rho \ast }M^{\mu
}{}_{\rho \sigma }\partial _{\mu }\psi ^{\sigma }-\frac{\mathrm{i}}{2}\left(
\partial _{\mu }\psi ^{\rho \ast }\right) M^{\mu }{}_{\rho \sigma }\psi
^{\sigma } \\
&&+\frac{\mathrm{i}}{2}\psi ^{\rho \ast }\left( M^{\mu }{}_{\rho \tau
}\Gamma ^{\tau }{}_{\sigma \mu }-\Gamma ^{\tau }{}_{\rho \mu }M^{\mu
}{}_{\tau \sigma }\right) \psi ^{\sigma } \\
&&+\frac{\mathrm{i}}{4}\omega _{\alpha \beta \mu }\psi ^{\rho \ast }\left[
M^{\mu }{}_{\rho \tau }\left( \mathbf{S}^{\alpha \beta }\right) ^{\tau
}{}_{\sigma }-\left( \mathbf{S}^{\alpha \beta \ast }\right) ^{\tau }{}_{\rho
}M^{\mu }{}_{\tau \sigma }\right] \psi ^{\sigma } \\
&&-\frac{m}{2}\left( \psi ^{\rho \ast }N_{\rho \sigma }\psi ^{\sigma \ast
}-\psi ^{\rho }N_{\rho \sigma }^{\ast }\psi ^{\sigma }\right) ,
\end{eqnarray*}%
from which it readily follows that%
\begin{eqnarray*}
\frac{\partial \mathcal{L}_{grav}^{W}}{\partial \psi ^{\rho \ast }} &=&\frac{%
\mathrm{i}}{2}M^{\mu }{}_{\rho \sigma }\partial _{\mu }\psi ^{\sigma
}-mN_{\rho \sigma }\psi ^{\sigma \ast } \\
&&+\frac{\mathrm{i}}{2}\left( M^{\mu }{}_{\rho \tau }\Gamma ^{\tau
}{}_{\sigma \mu }-\Gamma ^{\tau }{}_{\rho \mu }M^{\mu }{}_{\tau \sigma
}\right) \psi ^{\sigma }+\frac{\mathrm{i}}{4}\omega _{\alpha \beta \mu }%
\left[ M^{\mu }{}_{\rho \tau }\left( \mathbf{S}^{\alpha \beta }\right)
^{\tau }{}_{\sigma }-\left( \mathbf{S}^{\alpha \beta \ast }\right) ^{\tau
}{}_{\rho }M^{\mu }{}_{\tau \sigma }\right] \psi ^{\sigma }, \\
\frac{\partial \mathcal{L}_{grav}^{W}}{\partial \partial _{\mu }\psi ^{\rho
\ast }} &=&-\frac{\mathrm{i}}{2}M^{\mu }{}_{\rho \sigma }\psi ^{\sigma },
\end{eqnarray*}%
from which in turn it follows that the Euler-Lagrange equations of motion
for $\psi $ are given by%
\begin{eqnarray*}
0 &=&-\left[ \frac{1}{\sqrt{-g}}\partial _{\mu }\left( \sqrt{-g}\frac{%
\partial \mathcal{L}_{grav}^{W}}{\partial \partial _{\mu }\psi ^{\rho \ast }}%
\right) -\frac{\partial \mathcal{L}_{grav}^{W}}{\partial \psi ^{\rho \ast }}%
\right] \\
&=&\mathrm{i}M^{\mu }{}_{\rho \sigma }\partial _{\mu }\psi ^{\sigma
}-mN_{\rho \sigma }\psi ^{\sigma \ast } \\
&&+\frac{\mathrm{i}}{2}\left( \partial _{\mu }M^{\mu }{}_{\rho \sigma
}+\Gamma ^{\tau }{}_{\tau \mu }M^{\mu }{}_{\rho \sigma }+M^{\mu }{}_{\rho
\tau }\Gamma ^{\tau }{}_{\sigma \mu }-\Gamma ^{\tau }{}_{\rho \mu }M^{\mu
}{}_{\tau \sigma }\right) \psi ^{\sigma } \\
&&+\frac{\mathrm{i}}{4}\omega _{\alpha \beta \mu }\left[ M^{\mu }{}_{\rho
\tau }\left( \mathbf{S}^{\alpha \beta }\right) ^{\tau }{}_{\sigma }-\left( 
\mathbf{S}^{\alpha \beta \ast }\right) ^{\tau }{}_{\rho }M^{\mu }{}_{\tau
\sigma }\right] \psi ^{\sigma },
\end{eqnarray*}%
using the identity $\partial _{\mu }\sqrt{-g}=\sqrt{-g}\Gamma ^{\tau
}{}_{\tau \mu }$. Expressing $\partial _{\mu }M^{\mu }{}_{\rho \sigma }$ in
terms of $\nabla _{\mu }M^{\mu }{}_{\rho \sigma }$ (and three Christoffel
terms), these equations may be rewritten as%
\begin{eqnarray*}
0 &=&\mathrm{i}M^{\mu }{}_{\rho \sigma }\nabla _{\mu }\psi ^{\sigma
}-mN_{\rho \sigma }\psi ^{\sigma \ast } \\
&&+\frac{\mathrm{i}}{2}\left\{ \nabla _{\mu }M^{\mu }{}_{\rho \sigma }+\frac{%
1}{2}\omega _{\alpha \beta \mu }\left[ M^{\mu }{}_{\rho \tau }\left( \mathbf{%
S}^{\alpha \beta }\right) ^{\tau }{}_{\sigma }-\left( \mathbf{S}^{\alpha
\beta \ast }\right) ^{\tau }{}_{\rho }M^{\mu }{}_{\tau \sigma }\right]
\right\} \psi ^{\sigma }.
\end{eqnarray*}%
Being now manifestly tensorial, there being no explicit Christoffel symbols
present, these equations can be rewritten, by unproblematically
raising/lowering various indices, in matrix form as follows (note boldfaced
nabla, compare previous remark):%
\begin{eqnarray*}
\mathbf{0} &=&\mathrm{i}\mathbf{M}^{\mu }\mathbf{\nabla }_{\mu }\mathbf{\psi 
}-m\mathbf{N\psi }^{\ast }+\frac{\mathrm{i}}{2}\left[ \nabla _{\mu }\mathbf{M%
}^{\mu }+\frac{1}{2}\omega _{\alpha \beta \mu }\left( \mathbf{M}^{\mu }%
\mathbf{S}^{\alpha \beta }-\mathbf{S}^{\alpha \beta \hat{\dagger}}\mathbf{M}%
^{\mu }\right) \right] \mathbf{\psi } \\
&=&\mathrm{i}\mathbf{M}^{\mu }\mathbf{\nabla }_{\mu }\mathbf{\psi }-m\mathbf{%
N\psi }^{\ast }+\frac{\mathrm{i}}{2}\left\{ \nabla _{\mu }\mathbf{M}^{\mu }+%
\frac{1}{2}\omega _{\alpha \beta \mu }\left( 2\mathbf{M}^{\mu }\mathbf{S}%
^{\alpha \beta }+\left[ \mathbf{V}^{\alpha \beta },\mathbf{M}^{\rho }\right]
+\left( \mathbf{V}^{\alpha \beta }\right) ^{\rho }{}_{\sigma }\mathbf{M}%
^{\sigma }\right) \right\} \mathbf{\psi } \\
&=&\mathrm{i}\mathbf{M}^{\mu }\mathbf{D}_{\mu }\mathbf{\psi }-m\mathbf{N\psi 
}^{\ast }+\frac{\mathrm{i}}{2}\left( D_{\mu }\mathbf{M}^{\mu }\right) 
\mathbf{\psi } \\
&=&\mathrm{i}\mathbf{M}^{\mu }\mathbf{D}_{\mu }\mathbf{\psi }-m\mathbf{N\psi 
}^{\ast } \\
&\equiv &\mathbf{E}_{grav}^{W},
\end{eqnarray*}%
using the identity $\left( \mathbf{S}^{\alpha \beta \ast }\right) _{\rho
}{}^{\tau }=\left( \mathbf{S}^{\alpha \beta \hat{\dagger}}\right) ^{\tau
}{}_{\rho }$, and Eqs. (\ref{Eq:SMLorentz}) and (\ref{Eq:MCovDer}). But $%
\mathbf{E}_{grav}^{W}$, as thus defined, is simply $\mathbf{E}_{free}^{W}$,
Eq. (\ref{Eq:EFreeWorld}), subjected to the substitution Eq. (\ref%
{Eq:LorentzCovDerWorldMatrix}), in conjunction with letting $\mathbf{M}^{\mu
}$ and $\mathbf{N}$ become generically spacetime-dependent. Therefore, in
the present formalism, the Lorentz gauging procedure commute with the
Euler-Lagrange variational procedure, as asserted in the Introduction.

\section*{Conclusion}

The formalism, as presented in this paper, for the coupling of a spinor
field to the gravitational field has the following two distinguishing
properties:

\begin{itemize}
\item Commutativity of the Lorentz gauging procedure and the Euler-Lagrange
variational procedure, respectively, compare Sec. \ref{Sec:Action}. In
contrast, compare the Introduction, this property is \textit{not} satisfied
by the standard vierbein formalism for the coupling of a Dirac fermion to
the gravitational field, the equivalence principle thus seemingly being here
violated, quite unsettlingly.

\item The use of world indices only, there being present neither Lorentz
vector indices nor Lorentz spinor indices. Although this property is
regarded by the author to be much less important than the commutativity
property, previous bullet item, it nonetheless endows the formalism with the
following two attractive properties: 1.) concerning indices, it is more
parsimonious than the standard formalism is, and 2.) it seems to treat all
types of fields, tensorial and spinorial ones, on an equal footing, contrary
to the standard formalism.
\end{itemize}

\noindent There remains of course much to be investigated. Possible subjects
are the following ones, just to mention some:

\begin{itemize}
\item Internal symmetry of the Lagrangian, and the coupling of $\mathbf{\psi 
}$ to the electromagnetic field.

\item Discrete symmetries: $C$, $P$, and $T$.

\item Free-particle solutions of the equations of motion for $\mathbf{\psi }$%
.

\item Hamiltonian and the possible quantization of the theory.
\end{itemize}

\noindent The author intends to return to some or all of these issues in one
or more future papers while at the same time of course most warmly welcoming
anyone interested in contributing.

\section{Appendix}

The derivative $\mathbf{D}_{\mu }\mathbf{\psi }$, as defined by Eq. (\ref%
{Eq:LorentzCovDerWorldMatrix}), is a proper Lorentz covariant derivative in
the sense that it transforms as $\delta \left( \mathbf{D}_{\mu }\mathbf{\psi 
}\right) =\frac{1}{2}\left( d\theta _{\alpha \beta }\right) \mathbf{S}%
^{\alpha \beta }\mathbf{D}_{\mu }\mathbf{\psi }$ under Eq. (\ref%
{Eq:LorentzTransWorld}) with $d\theta _{a\beta }$ arbitrary. Proof: The
left-hand side is given by (note various boldface nablas)%
\begin{eqnarray*}
\delta \left( \mathbf{D}_{\mu }\mathbf{\psi }\right) &\equiv &\mathbf{D}%
_{\mu }\left( \delta \mathbf{\psi }\right) +\left( \delta \mathbf{D}_{\mu
}\right) \mathbf{\psi } \\
&=&\left( \mathbf{\nabla }_{\mu }+\frac{1}{2}\omega _{\alpha \beta \mu }%
\mathbf{S}^{\alpha \beta }\right) \left[ \frac{1}{2}\left( d\theta _{\gamma
\delta }\right) \mathbf{S}^{\gamma \delta }\mathbf{\psi }\right] +\frac{1}{2}%
\left( \delta \omega _{\alpha \beta \mu }\right) \mathbf{S}^{\alpha \beta }%
\mathbf{\psi } \\
&=&\frac{1}{2}\left( d\theta _{\alpha \beta }\right) \mathbf{S}^{\alpha
\beta }\mathbf{\nabla }_{\mu }\mathbf{\psi }+\frac{1}{2}\left( \nabla _{\mu
}d\theta _{\alpha \beta }\right) \mathbf{S}^{\alpha \beta }\mathbf{\psi }+%
\frac{1}{4}\left( d\theta _{\alpha \beta }\right) \omega _{\gamma \delta \mu
}\mathbf{S}^{\gamma \delta }\mathbf{S}^{\alpha \beta }\mathbf{\psi }+\frac{1%
}{2}\left( \delta \omega _{\alpha \beta \mu }\right) \mathbf{S}^{\alpha
\beta }\mathbf{\psi },
\end{eqnarray*}%
using $\delta \nabla _{\mu }=0$ and $\delta \mathbf{S}^{\alpha \beta }=0$
due to $\delta g_{\mu \nu }=0$, and $\mathbf{\nabla }_{\rho }\mathbf{S}%
^{\alpha \beta }=\mathbf{0}$ (boldface nabla) due to $\nabla _{\rho }g_{\mu
\nu }=0$; and the right-hand side is given by (note boldface nabla)%
\begin{equation*}
\frac{1}{2}\left( d\theta _{\alpha \beta }\right) \mathbf{S}^{\alpha \beta }%
\mathbf{D}_{\mu }\mathbf{\psi }=\frac{1}{2}\left( d\theta _{\alpha \beta
}\right) \mathbf{S}^{\alpha \beta }\left( \mathbf{\nabla }_{\mu }+\frac{1}{2}%
\omega _{\gamma \delta \mu }\mathbf{S}^{\gamma \delta }\right) \mathbf{\psi }%
.
\end{equation*}%
Using Eq. (\ref{Eq:LorentzSpinorRep}), these two expressions will be equal
if and only if%
\begin{eqnarray*}
\left( \delta \omega _{\alpha \beta \mu }\right) \mathbf{S}^{\alpha \beta }
&=&\frac{1}{2}\left( d\theta _{\alpha \beta }\right) \omega _{\gamma \delta
\mu }\left[ \mathbf{S}^{\alpha \beta },\mathbf{S}^{\gamma \delta }\right]
-\left( \nabla _{\mu }d\theta _{\alpha \beta }\right) \mathbf{S}^{\alpha
\beta } \\
&=&-\frac{1}{2}\left( d\theta _{\alpha \beta }\right) \omega _{\gamma \delta
\mu }\left( g^{\alpha \gamma }\mathbf{S}^{\beta \delta }-g^{\alpha \delta }%
\mathbf{S}^{\beta \gamma }-g^{\beta \gamma }\mathbf{S}^{\alpha \delta
}+g^{\beta \delta }\mathbf{S}^{\alpha \gamma }\right) -\left( \nabla _{\mu
}d\theta _{\alpha \beta }\right) \mathbf{S}^{\alpha \beta } \\
&=&-\left[ \left( d\theta ^{\gamma }{}_{\alpha }\right) \omega _{\gamma
\beta \mu }+\left( d\theta ^{\gamma }{}_{\beta }\right) \omega _{\alpha
\gamma \mu }+\nabla _{\mu }d\theta _{\alpha \beta }\right] \mathbf{S}%
^{\alpha \beta },
\end{eqnarray*}%
which is in fact so due to Eq. (\ref{Eq:LorentzTransSpinConnWorld}). End of
proof.

This proof is structurally analogous to the proof in the standard formalism
for the coupling of a Dirac spinor field to the gravitational field, compare
again \cite[Sec. 31.A]{Weinberg: QFT} and \cite[Sec. 12.1]{GSW}, that the
Lorentz covariant derivative of a Dirac spinor field transforms properly
under local Lorentz transformations. It is given here, nonetheless, for the
benefit of the reader.

\end{document}